## Ciencias Físicas

Artículo original

# Análisis de polaridades magnéticas en regiones activas para la predicción de fulguraciones solares

## Analysis of magnetic polarities in active regions for the prediction of solar flares

Natalia Granados-Hernández, Santiago Vargas-Domínguez

Grupo de Astrofísica Solar, Observatorio Astronómico Nacional, Universidad Nacional de Colombia, Bogotá, Colombia

## Resumen

Las regiones solares activas, y los procesos que en ellas ocurren, han sido estudiadas y analizadas ampliamente y se han elaborado muchos tipos de modelos y caracterizaciones de los distintos eventos eruptivos que tienen lugar en la atmósfera solar. Las regiones más características son aquellas que tienen polaridad magnética opuesta y que, en su mayoría, generan eventos explosivos, como las denominadas fulguraciones solares. Las fulguraciones son intensas explosiones que suceden en la atmósfera solar, que pueden llegar a tener efectos adversos sobre la Tierra y la tecnología desarrollada por el ser humano, además de ser determinantes en el llamado clima espacial, por lo cual se ha intentado predecir su aparición. En este estudio se desarrolló un modelo predictivo de fulguraciones solares de clase mayor a M5 con base en los propuestos por **Korsos, *et al.*** (2014, 2015), utilizando la relación existente entre las fulguraciones y las regiones activas bipolares. En el análisis se tuvieron en cuenta las áreas de las umbras de las manchas solares de polaridad opuesta, su campo magnético promedio y los baricentros de cada una de estas manchas en una muestra de tres regiones activas. Como resultado se generó un modelo predictivo encontrando la variación temporal de cantidades relacionadas con los baricentros magnéticos que se produce por la evolución de las manchas solares, con lo que se confirmaron resultados previos consignados en la literatura. Se hizo un análisis estadístico para inspeccionar si después de una fulguración, puede tener lugar otra en las horas siguientes.

**Palabras clave:** Actividad solar; Regiones activas bipolares; Fulguración; Manchas solares.

## Abstract

Solar active regions and the processes that occur in them have been extensively studied and analyzed and many types of models and characterizations have been proposed for the occurrence of different eruptive events that take place in the solar atmosphere. The most characteristic of these regions are those that have opposite magnetic polarity, which, in their majority, generate explosive events such as the so-called solar flares. The flares are intense explosions occurring in the solar atmosphere with adverse effects on the Earth and the technology developed by humans, and they are also determining factors in the so-called space weather. For this reason, attempts have been made to predict the occurrence of these events. In the present study, we developed a predictive model of solar flares higher than M5 based on the articles proposed by **Korsos, *et al.*** (2014, 2015) using the relationship between the flares and the bipolar active regions. The analysis took into account the areas of the sunspots' umbra of opposite polarity, their average magnetic field, and the magnetic barycenter from each sunspot in the region for a sample of three active regions to find the temporal variation due to the evolution of the sunspots, thus confirming previous results reported in the literature. We made a statistical analysis to determine whether after a flare occurs, another can arise in the subsequent hours.

## Introducción

Desde hace más de un siglo el estudio del Sol y sus regiones activas ha sido continuo debido a la gran importancia que tienen la aparición y la evolución de eventos explosivos como las denominadas fulguraciones solares. Estos fenómenos pueden causar ciertos efectos adversos en la Tierra, especialmente en la tecnología desarrollada por la especie humana en los últimos cincuenta años.

Las fulguraciones de tipo X pueden emitir radiación de alta energía hacia el exterior del Sol a altas velocidades. La luz que emiten alcanza nuestro planeta a los 8 minutos y 19 segundos del evento y las partículas que expulsa pueden colisionar con el campo magnético terrestre hasta aproximadamente tres días después de ocurrido el evento en la atmósfera solar y afectar la actividad y el comportamiento de la atmósfera de nuestro planeta.

Dependiendo de la intensidad de la fulguración, en la Tierra se pueden experimentar desde auroras boreales, sin consecuencias negativas, hasta efectos realmente contraproducentes. Por ejemplo, una gran explosión puede causar anormalidades en la ionósfera, las denominadas tormentas geomagnéticas, las cuales pueden perjudicar y afectar las comunicaciones y las posiciones satelitales a nivel global, poner en peligro a astronautas en misiones que estén orbitando la Tierra e, incluso, sobrecargar y generar apagones en redes de distribución eléctrica. Esto sucede debido a que las partículas expelidas en las fulguraciones llegan a la parte superior de la atmósfera y crean corrientes eléctricas fuertes allí, generando cambios drásticos en las propiedades electromagnéticas que posteriormente influyen en las redes eléctricas de la Tierra. Asimismo, pueden alterar las comunicaciones interrumpiendo la propagación de ondas de radio y de otras señales de las que nuestra sociedad actual depende en gran medida. Los más vulnerables frente a estos eventos son los satélites que orbitan la Tierra, así como las personas que se encuentren en estaciones espaciales, ya que esta radiación impacta directamente, sin el filtro que constituye la atmósfera (**Collins,** 2013). El fenómeno solar más intenso del que tenemos registro ocurrió a finales de agosto de 1859 y se le denominó "Evento Carrington"; se produjo por una fulguración acompañada de una gran liberación de material solar (eyección de masa coronal o *Coronal Mass Ejection*-CME), que generó una tormenta geomagnética de grandes proporciones en la Tierra, produciendo auroras incluso en latitudes cercanas al ecuador, como la reportada en Colombia (**Moreno-Cárdenas,** *et al.,* 2016).

Por esto es importante desarrollar modelos predictivos de fulguraciones (**Korsos,** *et al.,* 2014, 2015), en los cuales se usa la dependencia de las regiones activas magnéticamente bipolares con la formación de fulguraciones. Las variables de interés son las áreas y la magnitud media del campo magnético en las umbras de manchas solares en regiones de polaridad opuesta, y la distancia entre los baricentros de dichas zonas magnéticamente opuestas.

## Herramientas para el análisis de los magnetogramas

En el desarrollo del estudio se usaron magnetogramas adquiridos con el instrumento Michelson-Doppler Imager (MDI) a bordo del satélite Solar and Heliospheric Observator*y* (SOHO, **Domingo,** *et al.,* 1995), en el formato line-of-sight (LOS) magnetic field y tomados del catálogo Virtual Solar Observatory. Se seleccionó una muestra de regiones activas bipolares, es decir, zonas donde se observa una marcada diferencia a gran escala entre la polaridad positiva y la negativa del campo magnético.

### *Algoritmo*

Para el análisis de los magnetogramas se usó el programa Python (**Robitaille,** *et al.,* 2013). Las librerías requeridas fueron: matplotlib, para graficar los magnetogramas; astropy, con la que se puede manipular la información proporcionada por el magnetograma; numpy, con el cual se manipula el tamaño del arreglo y se encuentran algunas variables de interés; scipy para el manejo de algunas variables, y skimage, que permite manipular la resolución de la imagen.

Inicialmente se tomó el magnetograma y se lo transformó en un arreglo usando el paquete astropy.io.fits.getdata. Una vez transformado, se graficó con matplotlib para poder ubicar la región de interés. En la **figura 1** se muestra un ejemplo de uno de los magnetogramas del disco solar. Sobre la imagen del magnetograma se realizó una extracción de la región bipolar de interés (ver recuadro en la **figura 1**). La región recortada se destaca en la **figura 2**.

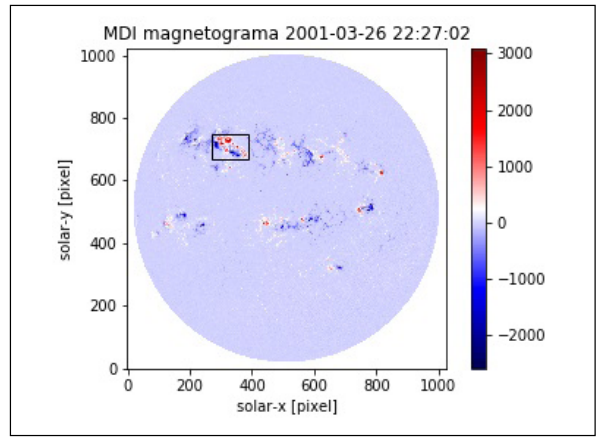

**Figura 1.** Magnetograma de la región activa NOAA AR 9393 del 26 de marzo de 2001, tomada con el satélite SOHO. La escala de color indica el valor del campo magnético en unidades de Gauss, con las polaridades positivas y negativas en rojo y azul. El recuadro que se observa en la figura destaca la región activa de interés.

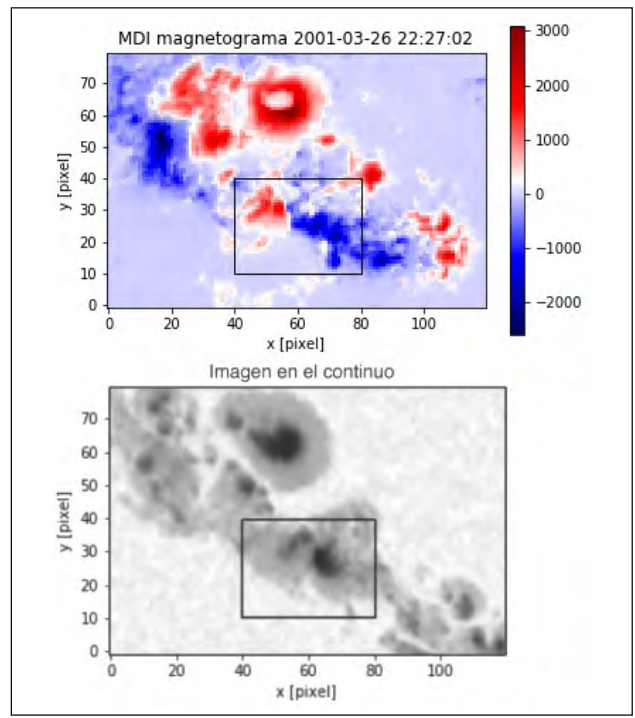

**Figura 2.** Región activa NOAA AR 9393 extraída de la figura 1. Panel superior: magnetograma. El recuadro destaca la subregión donde se encuentran las manchas solares con polaridad opuesta objeto del análisis. Panel inferior: imagen en el continuo en luz blanca de la región de interés tomada con el telescopio espacial SOHO.

Cuando se trata de una región solar activa compleja con varios pares de manchas solares, se debe seleccionar una región más pequeña y simple que incluya solo un par de manchas de polaridad opuesta, por lo cual se extrae una subregión sobre la cual se hace el análisis (ver recuadro en la **figura 2**).

Este par simple de manchas solares tiene la característica de tener una línea de inversión de polaridad (*polarity inversion line*, PIL), es decir, una zona neutra bien definida que las separa y que caracteriza la región bipolar, es decir, aquella en la cual finalmente se determinarán todas las variables. En esta imagen recortada se aprecia notablemente el pixelado. El tamaño del pixel correspondió a 1,98 arcosegundos tanto para el eje $x$ como para el $y$. Para trabajar más adecuadamente se aplicó un suavizado a la imagen reduciendo el tamaño de los pixeles a la décima parte, lo que permitió identificar de mejor manera los bordes de la región bajo análisis, en particular las umbras, es decir las zonas con valores extremos (positivos y negativos) de campo magnético.

Para hallar el área de las umbras se usó una máscara sobre el magnetograma elaborada con el módulo skimage.measure.find_contours, el cual permite hallar contornos sobre la imagen teniendo en cuenta el valor del pixel (en Gauss). Dado que a simple vista es complicado hallar un borde asociado con la umbra, se hizo un cambio en el mapa de colores de la región para resaltar e identificar bordes de valores en la imagen.

Los mapas de color seleccionados (set3 y tab20), que en adelante se denominan Map1 y Map2, permitieron dividir la imagen en 12 y 20 niveles, respectivamente, con el fin de analizar qué delimitación de la umbra se adecuaba mejor al modelo de **Korsos, *et al.*** (2014, 2015). Una vez establecido el nivel más cercano a la umbra, se graficó y se verificó su correspondencia, tal como se ilustra en la **figura 3**.

Ya definido el valor asociado con el borde de la umbra, se calculó el área sumando la cantidad de pixeles que envolvían el contorno y haciendo la conversión a arcosegundos con los datos de la cabecera del FITS asociado con el magnetograma. Posteriormente, se calculó el valor promedio del campo magnético dentro del área seleccionada para la umbra usando el comando count_nonzero de la librería numpy y la media.

Para hallar el baricentro de las regiones de polaridad opuesta se usó la librería scipy y su comando ndimage.center_of_mass, el cual permite usar los pixeles que están dentro del contorno previamente seleccionado. Se usaron los siguientes cuatro métodos para encontrar el baricentro de las manchas solares de polaridad opuesta.

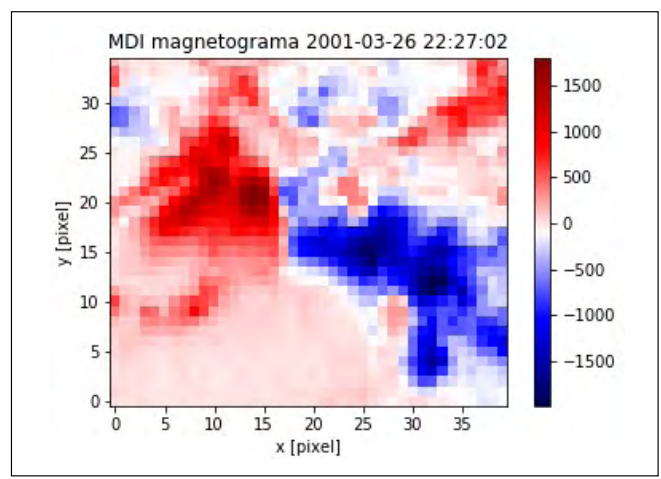

**Figura 3.** Magnetogramas de MDI de la región activa NOAA AR 9393 con contornos de magnitud del campo magnético. La columna de la izquierda muestra el magnetograma con los contornos de campo magnético en que se resalta la umbra de las manchas solares calculada con los mapas de contornos mostrados en la columna derecha con diferentes niveles correspondientes a Map1 (superior) y Map2 (inferior).

1. Una máscara que involucró los pixeles con un valor mayor o igual a 40 % del máximo y menor o igual al 40 % del mínimo de la región.
2. Una máscara que involucró los pixeles con un valor mayor o igual al 99 % del máximo y menor o igual al 99 % del mínimo de la región.
3. Una máscara que involucró los pixeles asociados dentro de la umbra encontrada a partir del mapa de colores Map1.
4. Una máscara que involucró los píxeles asociados dentro de la umbra encontrada a partir del mapa de colores Map2.
5. Un ejemplo de la ubicación de estos baricentros se ilustra en la **figura 4**, resaltando sus posiciones para ambas polaridades con símbolo de estrellas.

*Análisis de los datos*

Se analizaron magnetogramas tomados del instrumento Michelson-Doppler Imager (MDI) del Solar and Heliospheric Observatory (SOHO), con el fin de estudiar regiones activas bipolares con el método usado recientemente por varios autores (**Korsos,** *et al.,* 2014) para predecir un rango de tiempo en el que puedan generarse fulguraciones de categoría mayor a M5 y X.

Inicialmente se analizaron tres regiones activas usando el método ya descrito. Se calculó el área de las umbras de cada región, así como el valor medio del campo magnético en esta zona y la distancia entre los baricentros magnéticos de cada mancha de polaridad opuesta. Se ajustó el estudio de referencia del trabajo (**Korsos,** *et al.*, 2014) para determinar el campo magnético tomando los datos del campo magnético promedio dentro de la umbra. Posteriormente, se usó la definición de gradiente de campo magnético (*Magnetic Field Gradient,* $G_M$) y la generalización dada por **Korsos,** *et al.* (2015), así como el gradiente magnético horizontal ponderado (*Weighted Horizontal Magnetic Gradient*, $WG_M$). Estas cantidades se definen en las ecuaciones (1) y (2).

$$G_M = \left|\frac{B_p \cdot A_p - B_n \cdot A_n}{D}\right| \quad (1) \text{ y}$$

$$WG_M = \left|\frac{\Sigma_{p,i} B_{p,i} \cdot A_{p,i} - \Sigma_{n,j} B_{n,j} \cdot A_{n,j}}{D}\right| \quad (2),$$

donde $A$ y $B$ representan el área de las umbras y el campo magnético promedio dentro de estas, respectivamente, los índices $p$, $n$ denotan la polaridad positiva y negativa, y los índices $i$, $j$ la sucesión de las manchas usadas. Por último, $D$ es la distancia entre los baricentros de las manchas solares en la región activa de polaridad opuesta. Esta distancia se determina con la ecuación de la distancia entre dos puntos (3),

$$D = \sqrt{(x_p - x_n)^2 + (y_p - y_n)^2} \quad (3),$$

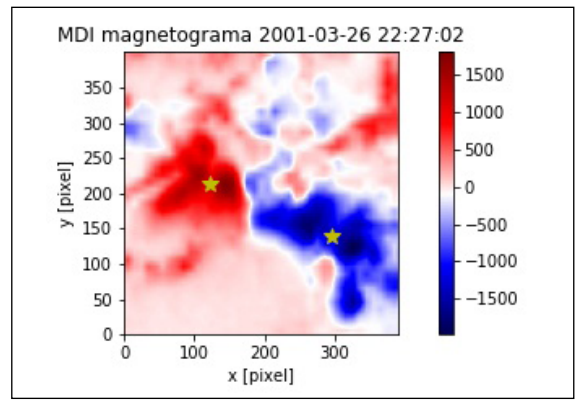

**Figura 4.** Magnetograma de la región activa de interés donde se marcan los baricentros magnéticos (con el símbolo de estrella) para las polaridades positivas y negativas (rojo y azul) hallados con el método 4

donde $(x_p, y_p)$ y $(x_n, y_n)$ son las coordenadas de los baricentros de las manchas con polaridad positiva y negativa, respectivamente. Esta ecuación se usa porque la distancia entre ellas es mucho más pequeña que el perímetro solar, por lo cual se puede aproximar a una distancia lineal.

Una vez se hallaron todas las variables mencionadas, se hicieron las gráficas de $G_M$ y $WG_M$ en el tiempo para conocer cuál de las dos variables se ajustaba mejor al modelo mencionado.

Las regiones activas que se analizaron fueron las siguientes:
- NOAA AR 8771, del 23 al 27 de noviembre de 1999
- NOAA AR 9393, del 26 de marzo al 2 de abril de 2001
- NOAA AR 10226, del 16 al 20 de diciembre de 2002

Para verificar la hora de aparición de estas fulguraciones se usó la hora de inicio de la base de datos Laboratory of X-Ray Astronomy of the Sun, LPI RAS, Russia, (https://tesis.lebedev.ru/en/).

Dado el modelo presentado en **Korsos, *et al.*** (2014, 2015) se esperaba encontrar un crecimiento de G_M y WG_M asociado con la acción de acercamiento de las manchas de polaridad opuesta y la creación de una hoja de corriente entre las líneas de su campo magnético. En una posterior fase de retroceso, estas variables decrecen cuando las manchas se separan, generando el proceso de reconexión magnética que produce la fulguración, como se ilustra en la **figura 5**.

Se estimó el tiempo entre el máximo de $G_M$ y $WG_M$ y la fulguración. Además, se aplicó un modelo estadístico para el cálculo del tiempo entre la primera fulguración y la posterior siguiendo el orden cronológico en cada región activa en caso de que en las siguientes no se tuviera un máximo asociado.

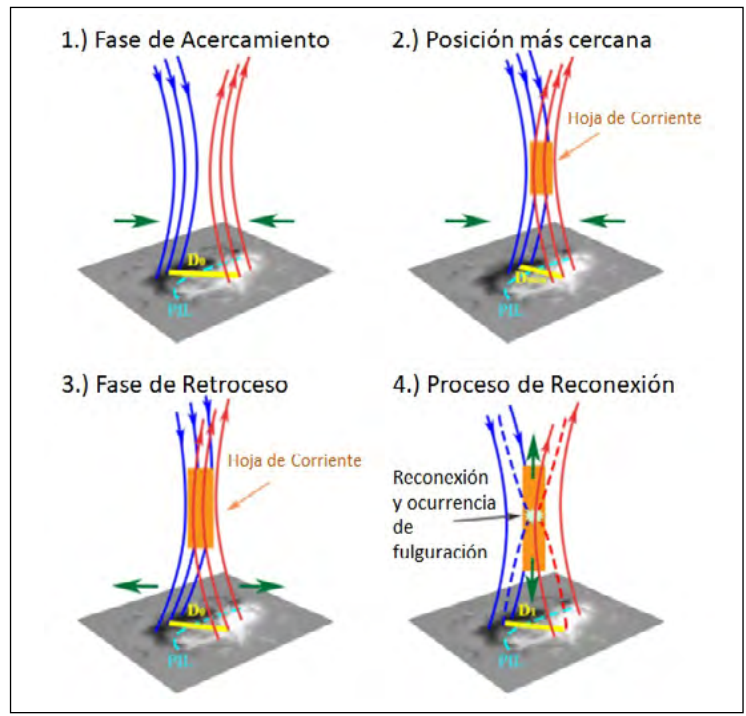

**Figura 5.** Proceso de formación de una fulguración asociado con el acercamiento (paneles 1 y 2) y alejamiento (paneles 3 y 4) de manchas de polaridad opuesta dentro de una región activa. Las líneas de color naranja y azul corresponden a las líneas magnéticas asociadas con las polaridades positiva y negativa, respectivamente. La línea punteada de color cian corresponde a la línea de inversión de polaridad (PIL) sobre el magnetograma. La línea amarilla denota la distancia D entre los baricentros de polaridad magnética. Fuente: adaptado de **Korsos, *et al.*** (2018)

## Resultados

Se encontró que el método más efectivo para hallar el tiempo estimado de aparición de una fulguración fue aquel en que se usó el gradiente magnético horizontal ponderado, $WG_M$, para todas las regiones activas de la muestra seleccionada y los métodos más efectivos para hallar los baricentros de las manchas solares de polaridad opuesta fue mediante los mapas de colores Map1 y Map2 (métodos 3 y 4).

### Región activa 8771

En esta región se encontró que habían ocurrido dos fulguraciones, la primera de clase M5.7, el 26 de noviembre de 1999 a las 01:10 UT, y la segunda de clase X1.4, el 27 de noviembre de 1999 a las 12:05 UT. Los valores calculados para esta región activa se encuentran en la **tabla 1**; se muestran los distintos máximos $WG_M^{max}$ para cada condición. El tiempo de aparición de la fulguración posterior al máximo en cada condición fue de 45 horas 59 minutos. En la **figura 6** (paneles superiores) se presentan las gráficas de evolución temporal de $WG_M$ donde se aprecia la dependencia del gradiente magnético horizontal ponderado con el tiempo. En las gráficas de la izquierda se refleja el análisis hecho con el mapa de colores Map1 y a la izquierda, con el mapa de colores Map2. En ambos casos se observó un crecimiento de $WG_M$ con un posterior decrecimiento lento asociado con la fase de retroceso que se da entre las manchas de polaridad opuesta.

**Tabla 1.** Fecha y máximo del gradiente magnético horizontal ponderado $WG_M^{max}$ previo a las fulguraciones M5.7 y X1.4 con su respectivo valor $WG_M^{flare}$ en el momento de su aparición

|  | Fecha | $WG_M^{max}$ [Wb/m] | $WG_M^{flare}$ M5.7 [Wb/m] | $WG_M^{flare}$ X1.4 [Wb/m] |
|---|---|---|---|---|
| $WG_{M,Map1}$ | 24/11/1999 3:11 | $8,02 \times 10^7$ | $5,61 \times 10^7$ | $7,59 \times 10^7$ |
| $WG_{M,Map2}$ | 24/11/1999 3:11 | $6,46 \times 10^7$ | $4,42 \times 10^7$ | $6,14 \times 10^7$ |

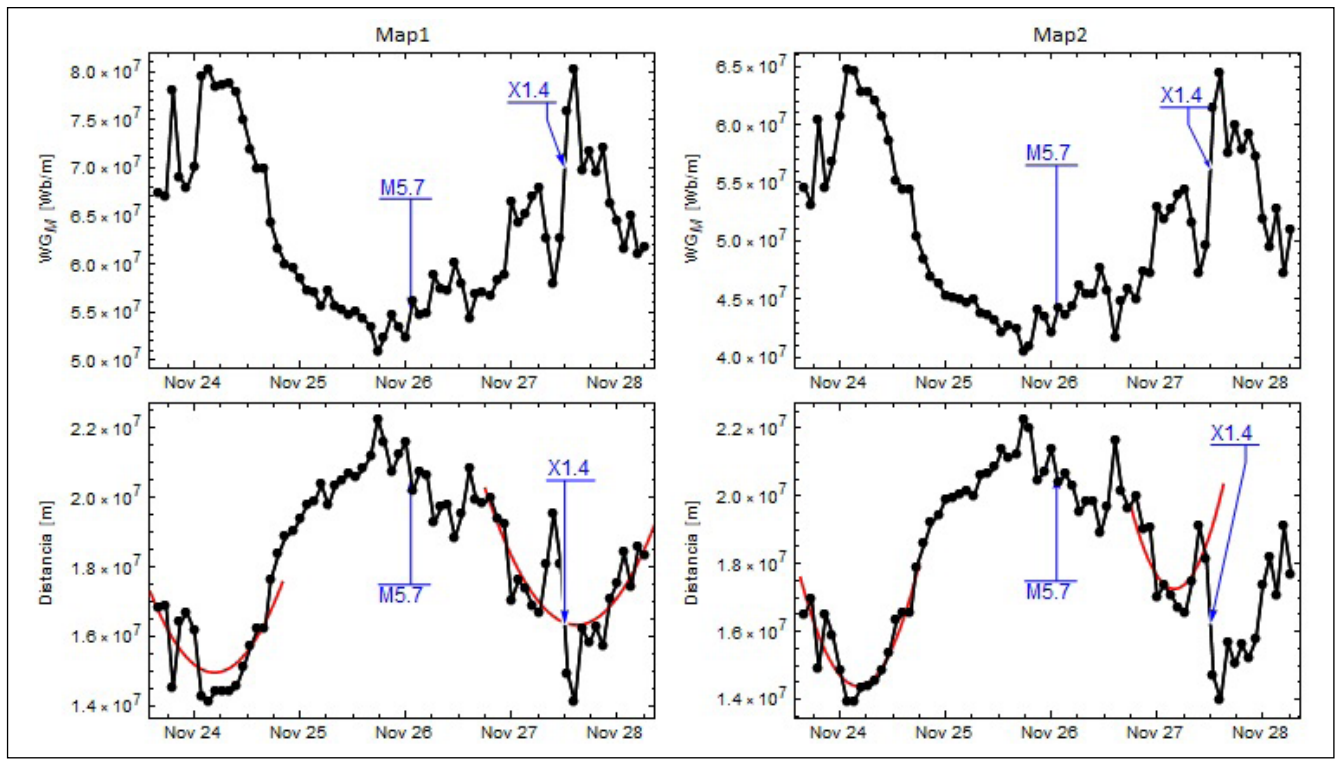

**Figura 6.** Variación temporal del gradiente magnético horizontal ponderado $WG_M$ para la región activa NOAA AR 8771. Columna izquierda: análisis para Map1. Columna derecha: análisis para Map2. Para cada condición en la parte inferior se encuentra la variación de la distancia de los baricentros de las umbras de polaridad magnéticamente opuestas. Se indican los momentos de aparición de fulguraciones de diferente tipo (en azul).

Las gráficas de comportamiento entre la distancia de los baricentros magnéticos (paneles inferiores de la **figura 6**) se hicieron análogamente y en estos se evidenció un comportamiento parabólico previo a la aparición de las fulguraciones que permite deducir que estas ocurren después de que los baricentros se separan notablemente, generando una reconexión entre las líneas de campo magnético que originan la fulguración.

### Región activa 9393

En esta región se encontró que habían ocurrido tres fulguraciones, la primera, de clase X1.7, el 29 de marzo de 2001 a las 09:57 UT, la segunda, de clase X1.4, el 2 de abril de 2001 a las 10:04 UT y la última, de clase X2.0, el 2 de abril de 2001 a las 21:32 UT.

Los datos encontrados para cada fulguración se encuentran en las **tablas 2, 3 y 4**. Por notación, los subíndices 1, 2 y 3 se relacionaron con las fechas y los máximos previos a cada una de las fulguraciones en orden cronológico.

En la **tabla 2** se muestra la fecha y el máximo $WG_{M,1}^{max}$ previo a la aparición de la primera fulguración de clase X1.7. El tiempo entre este máximo y la manifestación de la fulguración fue de 22 horas 45 minutos.

Por último, en la **tabla 4** se encuentran los máximos asociados con la última fulguración, de clase X20. El tiempo entre este máximo y la manifestación de la fulguración fue de 5 horas 32 minutos.

Se observó que se produjeron tres picos de $WG_M$ asociados con cada una de las fulguraciones (paneles superiores de la **figura 7**). En los dos casos esta variable creció hasta formarse un máximo de menor intensidad previo a la fulguración X1.7. Posteriormente, esta variable se mantuvo estable hasta el decrecimiento que produjo la fulguración X1.4, para luego crecer y disminuir rápidamente hasta generar una fulguración más intensa de tipo X20.

**Tabla 2.** Fecha y máximo del gradiente magnético horizontal ponderado $WG_{M,1}^{max}$ previo a la fulguración X1.7 con su respectivo valor $WG_{M,1}^{flare}$ en el momento de su aparición

|  | $Fecha_1$ [Wb/m] | $WG_{M,1}^{max}$ [Wb/m] | $WG_{M,1}^{flare}$ X1.7 [Wb/m] |
|---|---|---|---|
| $WG_{M,Map1}$ | 28/03/2001 11:12 | $1{,}73 \times 10^8$ | $1{,}58 \times 10^8$ |
| $WG_{M,Map2}$ | 28/03/2001 11:12 | $1{,}46 \times 10^8$ | $1{,}32 \times 10^8$ |

**Tabla 3.** Fecha y máximo del gradiente magnético horizontal ponderado $WG_{M,2}^{max}$ previo a la fulguración X1.4 con su respectivo valor $WG_{M,2}^{flare}$ en el momento de su aparición

|  | $Fecha_2$ [Wb/m] | $WG_{M,2}^{max}$ [Wb/m] | $WG_{M,2}^{flare}$ X1.4 [Wb/m] |
|---|---|---|---|
| $WG_{M,Map1}$ | 1/04/2001 11:12 | $1{,}77 \times 10^8$ | $1{,}46 \times 10^8$ |
| $WG_{M,Map2}$ | 1/04/2001 16:00 | $1{,}55 \times 10^8$ | $1{,}38 \times 10^8$ |

**Tabla 4.** Fecha y máximo del gradiente magnético horizontal ponderado $WG_{M,3}^{max}$ previo a la fulguración X20 con su respectivo valor $WG_{M,3}^{flare}$ en el momento de su aparición

|  | $Fecha_3$ [Wb/m] | $WG_{M,3}^{max}$ [Wb/m] | $WG_{M,3}^{flare}$ X20 [Wb/m] |
|---|---|---|---|
| $WG_{M,Map1}$ | 2/04/2001 16:00 | $1{,}90 \times 10^8$ | $1{,}47 \times 10^8$ |
| $WG_{M,Map2}$ | 2/04/2001 16:00 | $1{,}68 \times 10^8$ | $1{,}32 \times 10^8$ |

En ambos casos se puede apreciar que la evolución temporal de la distancia de los baricentros tuvo un comportamiento parabólico antes de la formación de las fulguraciones (paneles inferiores, **figura 7**).

### *Región activa 10226*

En esta región se observaron dos fulguraciones, la primera, de clase M2.4, que se produjo el 18 de diciembre de 2002 a las 06:31 UT, y la segunda, de clase M6.8, que se generó el 20 de diciembre de 2020 a las 13:13 UT.

En esta región se hizo una excepción y se tuvo en cuenta la fulguración de clase M2.4 porque en las gráficas de $WG_M$ (ver la figura del material suplementario, https://www.raccefyn.co/index.php/raccefyn/article/view/1196/2898) se pudo observar un pequeño pico posiblemente asociado con esta fulguración de menor clase. Sumado a ello, se consultó la base de datos de fulguraciones y esta fue la única de clase M producida ese día. En las **tablas 5 y 6** se presentan los datos obtenidos para las variables $WG_M$.

En la **tabla 5** se muestra la fecha y el máximo $WG_{M,1}^{max}$ anterior a la aparición de la primera fulguración de clase M2.4. El tiempo entre este máximo y la manifestación de la fulguración fue de 8 horas 7 minutos.

En la **tabla 6** se muestra la fecha y el máximo $WG_{M,2}^{max}$ previo a la aparición de la primera fulguración de clase M2.4. El tiempo entre este máximo y la manifestación de la fulguración fue de 22 minutos tanto para Map1 como Map2.

Se observó un comportamiento similar de los dos mapas con tiempos iguales entre el máximo y la fulguración en los dos casos (ver figura en material suplementario, https://www.raccefyn.co/index.php/raccefyn/article/view/1196/2898). Análogamente se observó un máximo inicial al que no le corres-pondió ninguna fulguración y que pudo deberse al uso de datos cercanos al limbo solar en donde hay efectos de proyección que afectan el cálculo del área de las regiones magnéticas.

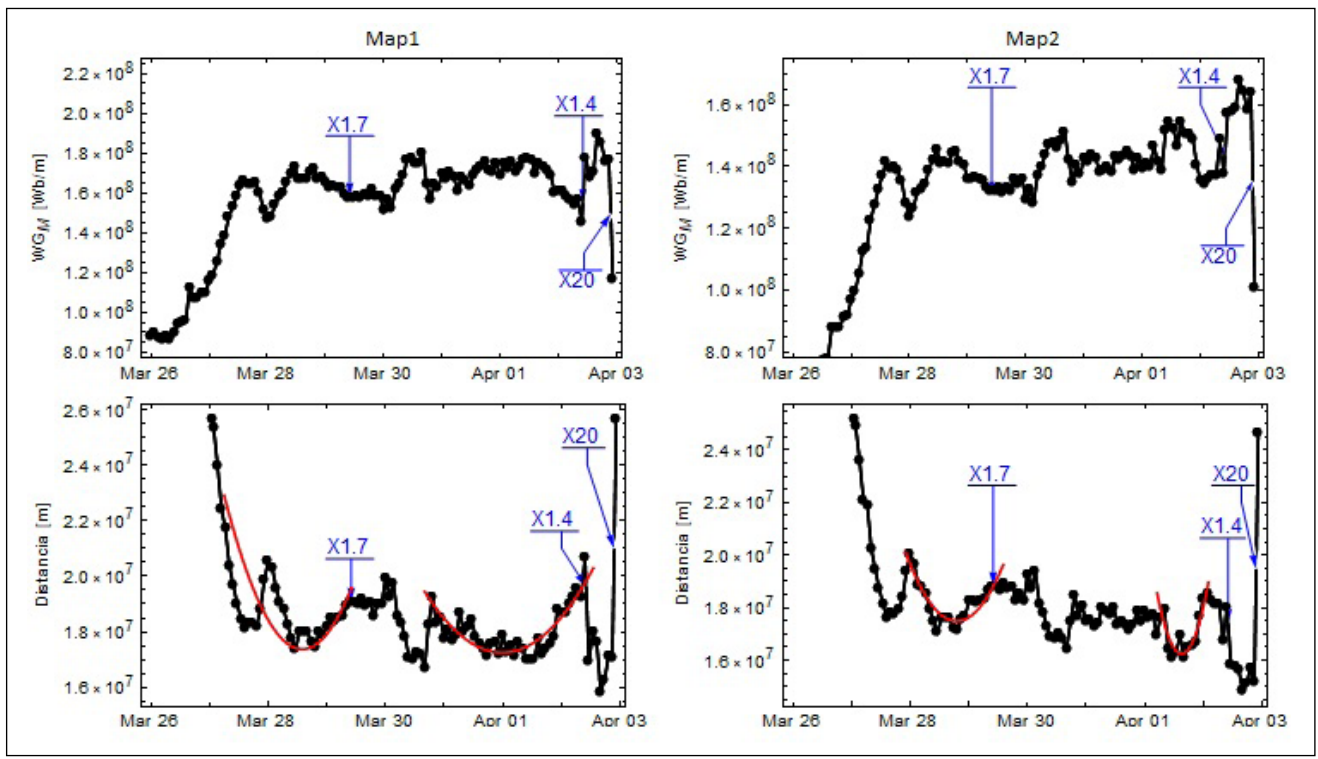

**Figura 7.** Variación temporal del gradiente magnético horizontal ponderado $WG_M$ para la región activa NOAA AR 9393, cuya descripción coincide con la expuesta en la figura 6.

**Tabla 5.** Fecha y máximo del gradiente magnético horizontal ponderado $WG_{M,1}^{max}$ previo a la fulguración M2.4 con su respectivo valor $WG_{M,1}^{flare}$ en el momento de su aparición

|  | $Fecha_1$ | $WG_{M,1}^{max}$ [Wb/m] | $WG_{M,1}^{flare}$ M2.4 [Wb/m] |
|---|---|---|---|
| $WG_{M,Map1}$ | 16/12/2002 17:36 | $1{,}99 \times 10^8$ | $6{,}84 \times 10^7$ |
| $WG_{M,Map2}$ | 17/12/2002 22:24 | $8{,}98 \times 10^7$ | $5{,}79 \times 10^7$ |

**Tabla 6.** Fecha y máximo del gradiente magnético horizontal ponderado $WG_{M,2}^{max}$ previo a la fulguración M6.8 con su respectivo valor $WG_{M,2}^{flare}$ en el momento de su aparición

|  | $Fecha_2$ [Wb/m] | $WG_{M,2}^{max}$ [Wb/m] | $WG_{M,2}^{flare}$ M6.8 [Wb/m] |
|---|---|---|---|
| $WG_{M,Map1}$ | 20/12/2002 12:51 | $3{,}61 \times 10^8$ | $3{,}14 \times 10^8$ |
| $WG_{M,Map2}$ | 20/12/2002 12:51 | $2{,}63 \times 10^8$ | $2{,}33 \times 10^8$ |

**Tabla 7.** Fecha y máximo del gradiente magnético horizontal ponderado $WG_{M,3}^{max}$ previo a la fulguración X20 con su respectivo valor $WG_{M,3}^{flare}$ en el momento de su aparición

| Fulguración | $WG_M^{\%}{,}_{Map1}$ | $WG_M^{\%}{,}_{Map2}$ |
|---|---|---|
| Región activa 8771 | | |
| M5.7 | 30,05 % | 31,57 % |
| X1.4 | 11,78 % | 12,86 % |
| Región activa 9393 | | |
| X1.7 | 8,67 % | 9,59 % |
| X1.4 | 17,51 % | 10,97 % |
| X20 | 22,63 % | 21,43 % |
| Región activa 10226 | | |
| M2.4 | 65,63 % | 35,52 % |
| M6.8 | 13,02 % | 11,42 % |

En el caso de la fulguración de clase M2.4 se observó un pico de menor intensidad con un decrecimiento que la generó, en tanto que para la fulguración de clase M6.8 hubo un aumento y una rápida disminución del $WG_M$, generando tiempos cortos entre el máximo y la fulguración, en este caso, de 22 minutos después del máximo.

En todos los casos se observó una tendencia parabólica en la distancia entre los baricentros de las manchas de polaridad opuesta previa a la formación de las fulguraciones.

### *Análisis estadístico de las regiones activas*

Se analizó la formación de las fulguraciones posteriores a una ocurrida previamente a partir del modelo estadístico usado en **Korsos,** *et al.* (2015, 2018) y se halló la diferencia porcentual $WG_M^{\%}$ entre el máximo de $WG_M^{max}$ y el valor del gradiente magnético en el momento de la fulguración $WG_M^{flare}$. En la **tabla 7** se presentan los porcentajes obtenidos para cada fulguración.

En el análisis se usó la base de datos del Laboratory of X-Ray Astronomy of the Sun, LPI, y se tuvieron en cuenta fulguraciones de todas las clases.

Se determinó que en todos los casos de fulguraciones, con excepción de la M2.4 de la región activa 10226, hubo otras en las 18 horas siguientes a su manifestación; por lo tanto, el método permitió estimar que: (i) para diferencias de porcentuales del gradiente magnético horizontal ponderado $WG_M^{\%}$ mayores al 35,5 % no son de esperar fulguraciones en las 18 horas posteriores; (ii) que si $WG_M^{\%}$ es menor a 30,1 % hay probabilidad de que en las siguientes 18 horas se genere otra fulguración.

## Conclusiones y discusión

En este estudio el objetivo era estudiar los modelos usados en **Korsos,** ***et al.*** (2014, 2015) a partir de magnetogramas tomados del observatorio satelital y el instrumento SOHO/MDI. Se usaron cuatro métodos para calcular los baricentros magnéticos de las manchas de polaridad opuesta para cada región activa y se constató que los óptimos para calcular el baricentro fueron los asociados con Map1 y Map2 (métodos 3 y 4). Estas condiciones generaron máximos $WG_M^{max}$ notables seguidos de fulguraciones en un rango menor a 45 horas.

En todas las gráficas asociadas con el modelo se observó el crecimiento y el decrecimiento de la variable de interés, lo que generó un máximo, ya que las manchas de polaridad magnética opuesta dentro de la región activa se acercaron e hicieron que las líneas equipotenciales de su campo se entrecruzaran formando una hoja de corriente. Posteriormente, en el momento en que se separaron ambas polaridades, se generó el proceso de reconexión magnética entre los campos magnéticos asociados con estas dos manchas, lo que desencadenó un evento de fulguración (**Figura 5**).

Se evidenció que cuando el crecimiento de $WG_M$ sucede rápidamente, hay probabilidad de que se genere una fulguración en las 10 horas posteriores al máximo. En algunas gráficas (**Figura 6**) se formaron máximos en sus extremos con la evolución temporal de $WG_M$, lo que puede deberse a que se tomaron datos cerca del limbo; en los estudios futuros se tendrá en cuenta este aspecto y se aplicarán correcciones para las regiones con efectos de borde.

Se hizo un análisis estadístico basado en las referencias de **Korsos,** ***et al***. (2015, 2018) y se encontró la diferencia porcentual de $WG_M^{\%}$ entre el máximo del gradiente magnético horizontal ponderado $WG_M^{max}$ y el valor de este al momento de manifestarse la fulguración $WG_M^{flare}$, con lo que se determinó el rango de tiempo en el que se pueden producir fulguraciones posteriores. Se obtuvo que cuando $WG_M^{\%}$ es mayor a 35,5 %, no es de esperarse la generación de fulguraciones en las siguientes 18 horas, en tanto que si es menor a 30,1 %, es dable esperar que se produzcan fulguraciones en las 18 horas posteriores. Si el valor de $WG_M^{\%}$ está entre 30,1 y 35,5 %, no se puede señalar certeramente la aparición de fulguraciones. Este análisis debe expandirse a una muestra más amplia de regiones activas para tener una base de datos con menor incertidumbre.

Se halló que la variable más importante en el momento de la producción de una fulguración fue la distancia entre los baricentros magnéticos de las manchas de polaridad opuesta, por lo tanto, se debe mejorar su método de obtención. Se espera complementar este trabajo con un algoritmo que permita analizar una muestra mucho más amplia de regiones activas con presencia de fulguraciones, para así mejorar el análisis estadístico y corroborar los resultados del método aquí presentados.

## Información suplementaria

**Figura suplementaria.** Variación temporal del gradiente magnético horizontal ponderado WGM para la región activa NOAA AR 10226. Columna izquierda: análisis para Map1. Columna derecha: análisis para Map2. Para cada condición en la parte inferior se encuentra la variación de la distancia de los baricentros de las umbras de polaridad magnéticamente opuestas. Se indican los momentos de aparición de fulguraciones de diferente tipo (en azul). Ver la figura suplementaria en https://www.raccefyn.co/index.php/raccefyn/article/view/1196/2898

## Contribución de los autores

NG: Descarga de datos, creación de rutinas de procesamiento, y visualización de los resultados; SVD: Apoyo en la creación de allgoritmos para el análisis de magnetogramas solares, análisis de los resultados e interpretación.

## Conflicto de intereses

Los autores declaran que no tienen ningún conflicto de intereses.

## Referencias